\documentclass[10pt]{article}
\usepackage{latexsym,graphicx}
\usepackage{latexsym,graphicx}
\newcommand{\be}{\begin{equation}}
\newcommand{\ee}{\end{equation}}
\def\n{\noindent}
\catcode `\@=11 \catcode `\@=12
\begin{document}
\begin{center}
\large{\bf {Redefining Heat and Work in the Right Perspective of 
Second-Law-of-Thermodynamics}} \\
\vspace{10mm}
\normalsize{R. C. Gupta $^1$,  Anirudh Pradhan $^2$ and Sushant Gupta $^3$} \\
\vspace{5mm}
\normalsize{$^{1}$ Institute of Technology \& Management (GLAITM), 
\\Mathura-281 406, India\\
E-mail: rcg\_iet@hotmail.com, rcgupta@glaitm.org}\\
\vspace{5mm} \normalsize{$^{2}$ Department of Mathematics, Hindu
Post-graduate College,\\Zamania-232 331, Ghazipur, India \\
E-mail: acpradhan@yahoo.com, apradhan@imsc.res.in}\\
\vspace{5mm} \normalsize{$^{3}$ Department of Physics,
University of Lucknow,\\Lucknow-226 007, India\\
E-mail: sushant1586@gmail.com}\\
\end{center}
\vspace{10mm}
\begin{abstract}
There are some misnomers and misconceptions about what is heat and
what is work; the recognition of heat and work is even more difficult
when it comes to categorize energy as heat or work. Since both heat
and work are energy the name-confusion does not make much difference
from engineering point of view, but re-defining `heat' and `work' in
the right-perspective of second-law-of-thermodynamics \cite {ref1}
is necessary to revise our understanding at fundamental level. It is
concluded that `heat is the energy carried by mass-less
\textit{photons} whereas work is energy carried by mass-ive material
\textit{fermions}'. Revised understanding of heat \& work in this
way has far reaching consequences in Physics [2-4]. The present
paper lays emphasis on re-defining heat and work, removing the
prevailing misconception, talks about single photon interaction and
heat property of photon. Also, interestingly, it is noted that
different fields of study such as `Thermodynamics' and `Relativity'
are interlinked.
\end{abstract}
\smallskip
\n Key words: Thermodynamics, Heat, Work, Energy, Irreversibility, 
Relativity, Photon, Photoelectric-effect, Compton-effect.\\
\section{Introduction}
Laws of thermodynamics are universally valid. The first-law is about the 
`\textit{conservation} of energy' whereas the second-law is about 
`\textit{conversion} of energy'. Engineers give equal weightage to both these
laws, but unfortunately the second-law has largely been ignored by physicists. 
The second-law of thermodynamics which basically tells about the `irreversibility' 
of the energy (heat and work) conversion process, has
far-reaching consequences [2-4]. Second law of thermodynamics \cite {ref1} implies 
that `although work can be fully converted to heat, but heat can-not be fully 
converted to work'. Efficiency of `work to heat conversion' could be $ = 100 \% $, 
but efficiency of `heat to work conversion' must be $ < 100\%$. The second law of
thermodynamics can be used as best criterion to judge whether a certain form of 
energy is work or is heat. Judging from this angle, as it is mentioned \cite {ref2} 
and elaborated in this paper that: it is very unfortunate that there are some 
misnomer and misconceptions which exists \& prevails specially for the heat,
right from the very beginning till now in our understanding, in books and literature. 
Though these misnomer and misconceptions do not make much difference from engineering 
point of view, re-defining `heat' and `work' in the right-perspective of 
second-law-of-thermodynamics is necessary to revise our understanding at fundamental 
level as it has far reaching consequences at deeper level in Physics [2-4].\\\\
It is further shown that: though heat is considered as a statistical
(bulk) property/aspect, but thermodynamics-laws are equally
applicable even for a single photon interaction such as
photoelectric-effect and Compton-effect. It is also found that widely
differing fields of study `thermodynamics' and `special-relativity'
are inter-supportive to each other [5-12].

\section{Misnomers \& Misconceptions about Heat \& Work
and Re-defining Energy as either Work or as Heat in view of
Second-Law of Thermodynamics}
In thermodynamic-processes `Heat' and `work' are generally
considered obvious, but there are some basic misconceptions too. The
so-called `heat of a hot-body', as per second law of thermodynamics,
is in fact not `heat' but `work' as it is due to vibration/motion of
atoms/molecules. In electronic and other processes where usually
`energy'- transfer/transition/conversion takes place, recognition of
heat \& work is even more difficult. \textit{\textbf{What is energy
? Is energy `heat' or `work' ? }} Identification of different types
of energies either as work or as heat is discussed in the next paragraphs 
and represented in Figure 1.\\\\
First consider the potential-energy, it is the work-done against a
force and is stored as potential energy; so potential energy is
`work'. It is well known that when a stone falls from height, the
potential energy changes to kinetic-energy, hence kinetic energy too
is `work' i.e., energy of motion and vibration of molecules and atoms
are work. Stored energy of an electron in atom is sum of its
potential energy and kinetic energy, is thus again the stored energy
is `work'. With furthermore arguments, it can be shown that: all
stored energies such as electrostatic energy, chemical energy,
internal energy, nuclear energy/mass energy $ mc^{2}$ etc. are
in-fact `work'. Though all stored energies are `work'; the energy
with moving particle is `work' if the energy is kinetic-energy and
it is `heat' if the energy is radiation-energy. In-fact
\textit{\textbf{`heat' is the energy carried by mass-less particle
such as photon, whereas all other form of energy carried by mass-ive
material particle are `work'}} (Figure 1). In other words,
messenger-particles bosons (photons) carry the `heat' as radiation,
whereas material-particles fermions (or fermion-groups as
atoms/molecules) carry `work' as kinetic \& stored energy in the
particle/matter. To illustrate - which energy is `work' and which
one is `heat' a ray-table is shown as follows (Figure 1). The
conclusion that `heat is carried by massless-particle
`\textit{photon}' reminds/revives the old caloric-concept [13-15] of
heat as `energy in transit' by
massless-fluid `\textit{caloric}'.\\\\
\begin{figure}[htbp]
\includegraphics[width=16cm,height=15cm,angle=0]{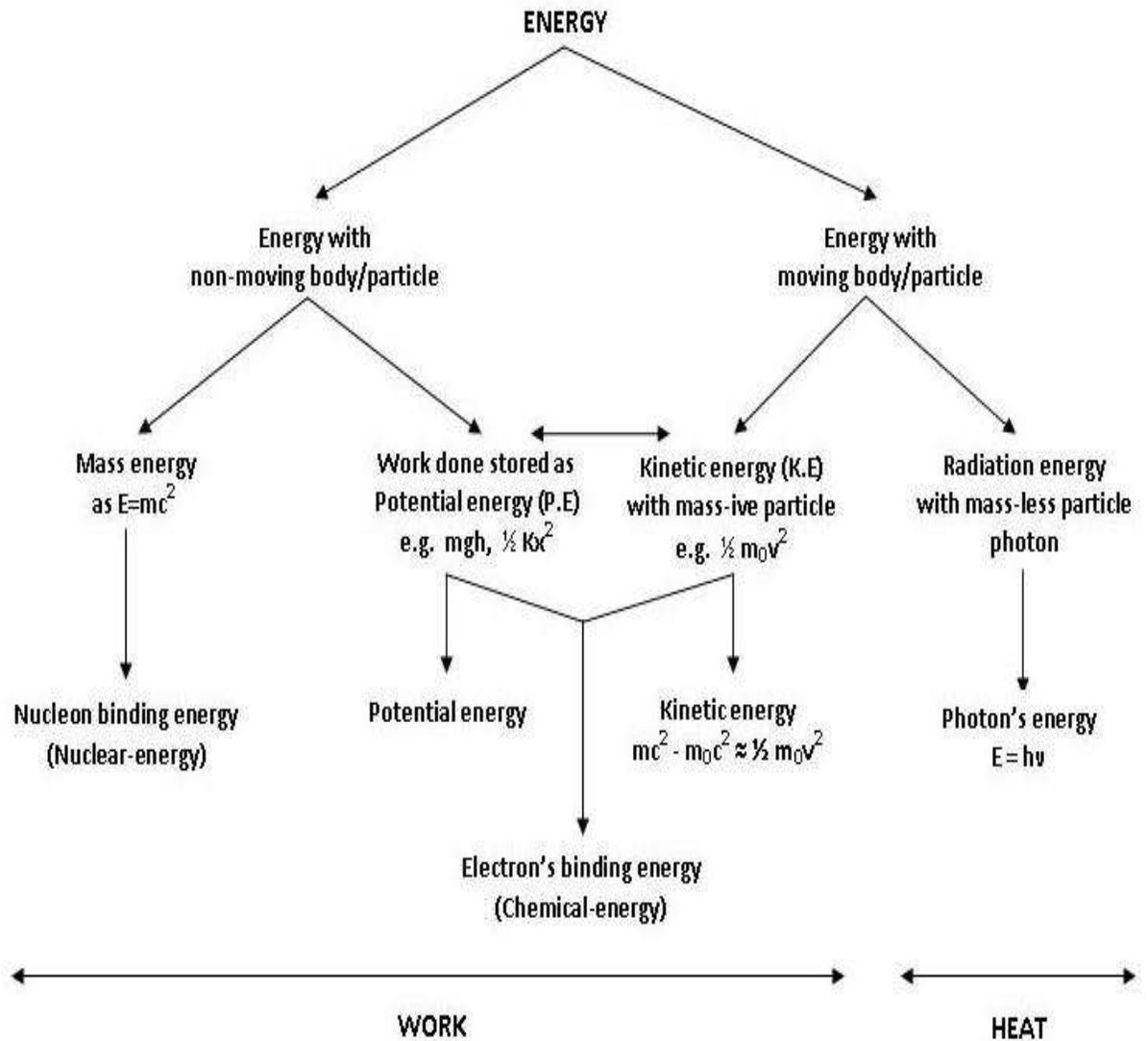}
\textbf{\caption{\textit{\textsl{`What is Energy?': In fact all
forms of energies are `work' except the radiation-energy which is
`heat'.}}}}
\end{figure}Let us look at the question `what is work' from a different 
angle. Work is the energy spent against a force, and this is either stored
as potential-energy or delivered as kinetic-energy. Kinetic-energy
is a `kinetic' manifestation of the `static' potential-energy, and
both are interchangeable to each other. Next question may be `what
are the forces against which work could be done'. There are four
fundamental forces of Nature : Gravitational, Electromagnetic and
Nuclear (Weak \& Strong). The other forces such as muscle-force,
hammer-force, explosive-force etc. are all basically electromagnetic
force. So, \textit{\textbf{all the energies, except
radiation-energy, are `work'}}.\\\\
Now let us examine what we mean by `heat of a hot body'. The word 
`heat' here is a misnomer, `\textit{heat} of the hot body' is due 
to motion and vibration of molecules and atoms and thus what we say 
as `heat' is actually due to kinetic and internal energy which are 
really in the category of `work'.\\\\
In thermodynamics, conventionally it is said that disordered motion
is heat whereas ordered motion is work, this too is a misconception.
Both disordered and ordered motion energy are work, we may say it
disordered work and ordered work. We cannot distinguish energy of a
crowd as heat and energy of a military-platoon as work; similarly,
we cannot distinguish that energy of ordinary light as heat and
energy of laser beam as work. The energy of crowd or platoon both
are work, whereas energy of light or laser both are heat; since in
the former example energy is carried by mass-ive material body
whereas in the later example energy is carried by mass-less particle
(photon).\\\\
Now let us consider the conventionally called `heat-transfer' 
\cite {ref16}; it is said that the transfer is
through (i) Conduction, (ii) Convection and (iii) Radiation. But a careful
re-examination of the fact would reveal that the literally real
`heat-transfer' is only through radiation only where the heat
(radiation energy) is transferred from one place to the other. The
transfer of energy in conduction and convection in in-fact the
kinetic-energy transfer from one atom/molecule to the other, so it
should be called `work-transfer'. Thus in place of `heat-transfer'
name, `energy-transfer' name would be a better name because this
includes all the three modes of energy transfer through conduction,
convection and radiation; `heat and mass transfer' name is also
acceptable because mass in itself implies energy too.\\\\
Now let us consider how energy is emitted and is absorbed. Energy
emission can be viewed as the release of energy when say
electron/atom/molecule goes from a higher energy level to a lower
energy level, releasing full energy change $ \triangle E$ (work) in
full to heat $ (h\nu)$. However, when heat $ (h\nu)$ falls on an
opaque body, a part of it may be reflected from surface-atoms and
the remaining is absorbed. This remaining energy goes to impart
kinetic energy \& internal energy to molecules \& atoms; only part
of `the absorbed part of heat' converts into `work', some energy
must go as waste (reflected/radiated from inside atoms). So, as per
second law of thermodynamics even the blackest of black-body can not
have $100\%$ absorptivity of heat i.e., $ \alpha $ can-not be equal
to 1 but must be very slightly less than 1.\\\\
When it comes to the meaning of temperature of a hot body;
conventionally it is thought that it is a measure of the level of
heat but truly speaking it is measure of level of energy (work)
contained in it. Also, we speak that internal energy is due to
vibration of atoms/molecules viz. $ \frac {1}{2}  m_{0}v^{2} = 3 kT
$ for solid \cite {ref17}, which means that temperature (T) is a
measure of kinetic energy (work).\\\\
Energy-wise heat (Q) and work (W) are equivalent by the Joule's
relationship $ W = J.Q $ . Though W \& Q are equivalent by the above
relation, but the transformation-processes of `work to heat' and
`heat to work' are different (efficiency-wise) from the point of
view of second law of thermodynamics. First law of thermodynamics
(energy conservation) states \textit{equivalence of W} \& \textit{Q
energy-wise}, whereas the second law of thermodynamics (entropy
increase) implies the \textit{non-equivalence or irreversibility in
its conversion}.
\section{Single Photon Interactions (such as Photoelectric-Effect
\& Compton-Effect) and Heat Property of the Photon} Conventionally,
heat is considered to be an averaged quantity, so questions arise
`whether a single photon has heat property and that whether applying
second-law-of-thermodynamics to single photon interaction such as
Photoelectric-effect and Compton-effect are reasonable or not?' The
authors answer both these questions as yes and yes, and firmly states
that indeed it is reasonable because: (i) Average of a single data
is also a quantity i.e., the data itself; a photon of energy $
(h\nu)$ has its average energy too as $ (h\nu)$, (ii) Though from
the very beginning (much before the birth of photon-concept) it has
been considered (in view of kinetic theory of gases) that heat is
statistical (bulk) property; but no-one established this or no-law
dictates that the second-law-of-thermodynamics can not be applied to
a single photon interaction. This means that nothing
forbids/prohibits and is therefore permissible; in fact single
photon interactions (such as Photoelectric and Compton effects) are
neat examples of applicability of both the first and second law of
thermodynamics, (iii) It is shown \cite {ref2} (and will be
reproduced again for clarity \& completeness) that not only the
applicability holds good very well but also shows new light
(understanding) i.e., it leads to conclusion that the
second-law-of-thermodynamics is in accordance with the basic idea of
special-relativity, (iv) The theory of relativity is applicable well
to single-particle therefore the second law of thermodynamics too
should be applicable, since both are shown to be interlinked [5-12].
\subsection{Photoelectric Effect} The photoelectric-effect
\cite{ref17} equation for a single photon interaction $ h\nu -
h\nu_{0} = \frac{1}{2} m_{0}v^{2} $ is indeed in accordance with the
second law of thermodynamics $ Q_{1} - Q_{2} = W $, which means that
the work-function $ (h\nu_{0}) $ corresponding to $ Q_{2} $ can
never be zero (and that is true). The thermodynamic-efficiency of
the process is therefore $ \eta = \frac{(\nu-\nu_{0})}{\nu}< 1 $.
\subsection{Compton Effect} The Compton-effect \cite {ref17} equation
for energy conversion (first law of thermodynamics) $ h\nu -
h\nu_{0} = \frac{1}{2} m_{0}v^{2}
$ or more precisely (with relativistic consideration)\\\\
$ h\nu - h\nu^{/} =
\frac{m_{0}c^{2}}{(1-\frac{v^{2}}{c^{2}})^{\frac{1}{2}}} -
m_{0}c^{2}
$\\\\
is also in accordance with the second law of thermodynamics $ Q_{1}
- Q_{2} = W $, which means that $ \nu^{/} $ can never be zero (and
that is true). The thermodynamic-efficiency of the process is,
similar to that of photoelectric effect, therefore $ \eta =
\frac{(\nu-\nu^{/})}{\nu}< 1 $.

\subsection{Temperature of a Photon} For a rough estimate of
temperature of a photon coming out from its source, the photon's
energy $(\frac{hc}{\lambda})$ is equated to the internal energy 3kT
of the source. This yields $ T = \frac{hc}{(3k \lambda)} $ which is
quite similar to the source-temperature estimate by Wein's law
\cite{ref17} $ T = \frac{hc}{(4.96k \lambda)} $. However, the single
photon temperature is usually high, but its effect is not
appreciably felt physically unless a large number of photons fall
upon and absorbed fully. Moreover, only a part of photon's energy is
absorbed on the receiver body and that too
further radiates-back as high-wavelength radiation.\\\\
Assigning a temperature to photon(s) is not something that is
completely new: cosmologist often ascribe a temperature of $
2.7^{0}K $ to Cosmic-Microwave-Background (CMB) radiation
(photons)[18-22]. Usually a `color' is linked to `temperature', and
whole range of VIBGYOR spectrum are linked to its corresponding
temperatures.\\\\
Photon do has heat property; heat of thermal-photon can be
physically felt, it is the photons which heats the food in the
microwave-oven or solar-cooker. Our sensory-organs too are made of
materials-atoms (fermions), part of the photon's energy is first
taken-up by the organ and the energy-transfer communicated to the
brain-matter. Even the instruments (material) absorb the energy and
the expansion (of say, Hg material) calibrated. Though photon's heat
is measured with the intervention of some material (sensory organ or
measuring-instrument), but it is true that photon do has heat
property and has a temperature too. If we are unable to see
something, that doesn't mean that it doesn't exist; we do not see
atoms but atoms do exist, and we have indirect evidences for its
presence. Ironically, everything which we see around us is visible
only due to presence of apparently-invisible photon. Presence of
photon and its heat, however, can be sensed as solar-radiation,
viewed as chemical-reaction taking place on photographic-plate or
can be heard as radio-song and so on. Even the absence-of-photon as
shadow indicates its existence. Photon is essential for
photosynthesis, and it can be further argued that the essence of
life is photon and its heat. \textit{Whole range of electromagnetic
radiation is heat; though our-sensory organs (hearing, seeing and
feeling-heat) may be more receptive only to certain range of
frequencies}.

\section{Thermodynamics and Relativity Linked through Photon}
Consider the Compton-effect again. `The outgoing photon ought to
exist, i.e., $ \nu^{/} $ can never be zero (which is in accordance
with the second law of thermodynamics)' is re-studied further
considering the two possibilities:\\\\
(i) If the incident photon $ (\nu) $ strikes the electron in
x-direction and if after collision the electron is deflected-away
from x-direction, the changed-photon $(\nu^{/}>0)$ ought to
come-out to balance the `momentum of electron in y-direction'.\\\\
(ii) If the incident photon $ (\nu) $ strikes the electron in
x-direction and if after collision the electron too moves in
x-direction, the changed-photon $(\nu^{/} > 0)$  after impact may
come-out in x-direction. This possibility is further analyzed as
follows.\\\\
The energy-equation of Compton-effect can be re-written using $
\nu^{/} > 0 $ as follows,\\\\
$ h\nu > \frac{m_{0}c^{2}}{(1-\frac{v^{2}}{c^{2}})^{\frac{1}{2}}} -
m_{0}c^{2} $\\\\
and the conservation of momentum for possibility (ii) using $
\nu^{/} > 0$ yields,\\\\
$ \frac{h\nu}{c} > \frac{m_{0}}{(1-\frac{v^{2}}{c^{2}})^{\frac{1}{2}}} v $\\\\
Putting the value of $ h\nu $ from one equation into the other and
after simplification, the following interesting result is found to
emerge-out as\\\\
$ 0 < v < c $\\\\
which is well \textit{in-accordance} with the basic concept of the
theory of Relativity.\\\\
This means that the result of second law of thermodynamics ( $
\nu^{/} $ is never zero, `heat to work' conversion $ \eta =
\frac{Q_{1}-Q_{2}}{Q_{1}} = \frac{W}{Q_{1}}< 1 $ ) is compatible
with the essence of special-relativity ( $ v < c $ i.e., no matter
how energetic may be the incident photon, velocity of electron can
not exceed velocity of light c, implying $ \beta = \frac{v}{c} < 1
$). It is amazing that how two quite different fields of study -
`Thermodynamics $(\eta < 1)$' and `Relativity $(\beta < 1)'$ are
inter-supportive and inter-linked to each other and appear to be the
two faces of the same coin. The link of Thermodynamics and
Relativity has been reported [5-12] earlier also.
\section{Heat and Work Revisited}
Probably the root-cause of the confusion about heat \& work arises
due to the fact that both (heat \& work) can give rise to a feel of
temperature to our senses. But if we widen our thought-horizon to
encompass both heat and work as ingredients (as follows) for
temperature, the confusion about heat and work diminishes. The
confusion/misconception however can not be removed completely in one
go, it needs time for our mind-set to adopt for the change/revision.
Work and heat can be considered to be related to
temperature (of say, solid) as follows:\\
\halign{ ~#\hfil& \quad\vtop{\hsize=0.5pc\noindent #\strut}&
\quad\hfil#& \quad\vtop{\hsize=7.3pc\noindent #\strut}& \quad\hfil#&
\quad\vtop{\hsize=10.0pc\noindent #\strut}\cr \noalign
{\smallskip\smallskip}
\\   For \textbf{work}&  \\&ingredient as
                       \\& internal energy
                        \\&$ \frac{1}{2} m_{0} v^{2} \approx 3kT $, \\
&$ T \approx \frac{m_{0} v^{2}}{6k} $ \cr
\noalign {\smallskip\smallskip}
\\   For \textbf{heat}&  \\&  ingredient as
                       \\ & radiation energy
                        \\&$ h \nu \approx 3kT $, \\&$ 
T \approx \frac{h \nu}{3k} $ \cr
\noalign {\smallskip\smallskip}} Now, let us re-view heat and work in
the popular equation of first law of thermodynamics (energy
conservation) : $ dQ = dU + dW + Losses = dU + dW + dL_{W} + dL_{Q}
$. Each terms are discussed in some details for clarification, as
follows.\\\\
\textbf{dQ}: It is the input-energy (heat) to the system. It is
usually the stored-energy (work) of the fuel which is released as
heat $ (h\nu) $ after ignition and which is then transferred to
atoms/molecules as kinetic-energy (work); part of this
kinetic-energy $ (\frac{1}{2}m_{0}v^{2}) $ goes as internal-energy
(dU) to the gas \& the container-body and part of which is used up
to produce work (dW) through piston-motion. Some input energy goes
as waste as radiation $ (dL_{Q}) $ as a necessity dictated by the 
second law of thermodynamics.\\\\
\textbf{dU}: It is the kinetic-energy $ (\frac{1}{2}m_{0}v^{2}) $ of
the atoms \& molecules of the exhaust gas \& the container body. It
is also called as internal or thermal-energy. As discussed (\&
tabulated) earlier, the kinetic-energy is like work but it gives a
feel of temperature, so usually misunderstood as heat.\\\\
\textbf{dW}: Part of the kinetic-energy of the combustion gases
produces useful work $ (dW = p.dV) $ through piston motion. Part of
this work is used for useful work (such as raising a load or moving
a vehicle); whereas part of it goes as waste against, say,
friction-resistance and goes as thermal-energy $ (dL_{W}) $ and
finally goes off as radiation heat.\\ 

For constant-volume process or for heating solid $ dV=0 $ hence for 
it $ dQ = dU + dL_{Q} $,similar to the second-law-of-thermodynamics equation $
Q_{1} =  W + Q_{2} $, indicating that dU is like work.\\\\
\textbf{Losses}: There are two types of losses viz., (1) $ dL_{W} $
and (2) $ dL_{Q} $, re-explained as follows.\\\\
(i) $ dL_{W} $ is the loss from the total work produced; the total 
work produced $ dW_{t} $ is equal to sum of
useful work $ (dW) $ and loss due to friction etc. $ (dL_{W}) $. Thus as per 
\textit{`first law'} of thermodynamics $ dQ =
dU + dW + dL_{W} $.\\\\
(ii) $ dL_{Q} $ is the loss dictated by the \textit{`second law'} of
thermodynamics, some energy must go as waste radiation
energy.\\\\
(iii) Taking into, account both the \textit{`first law'}
and \textit{`second law'} of thermodynamics, the final equation is
as follows: $ dQ = dU + dW + dL_{W} + dL_{Q} $.

\section{Another Look on Question of `What is Energy'}
Let us have another look (Figure 2) on the basic question - `what is
energy, is it heat or work?'.
\begin{figure}[htbp]
\centering
\includegraphics[width=16cm,height=19cm,angle=0]{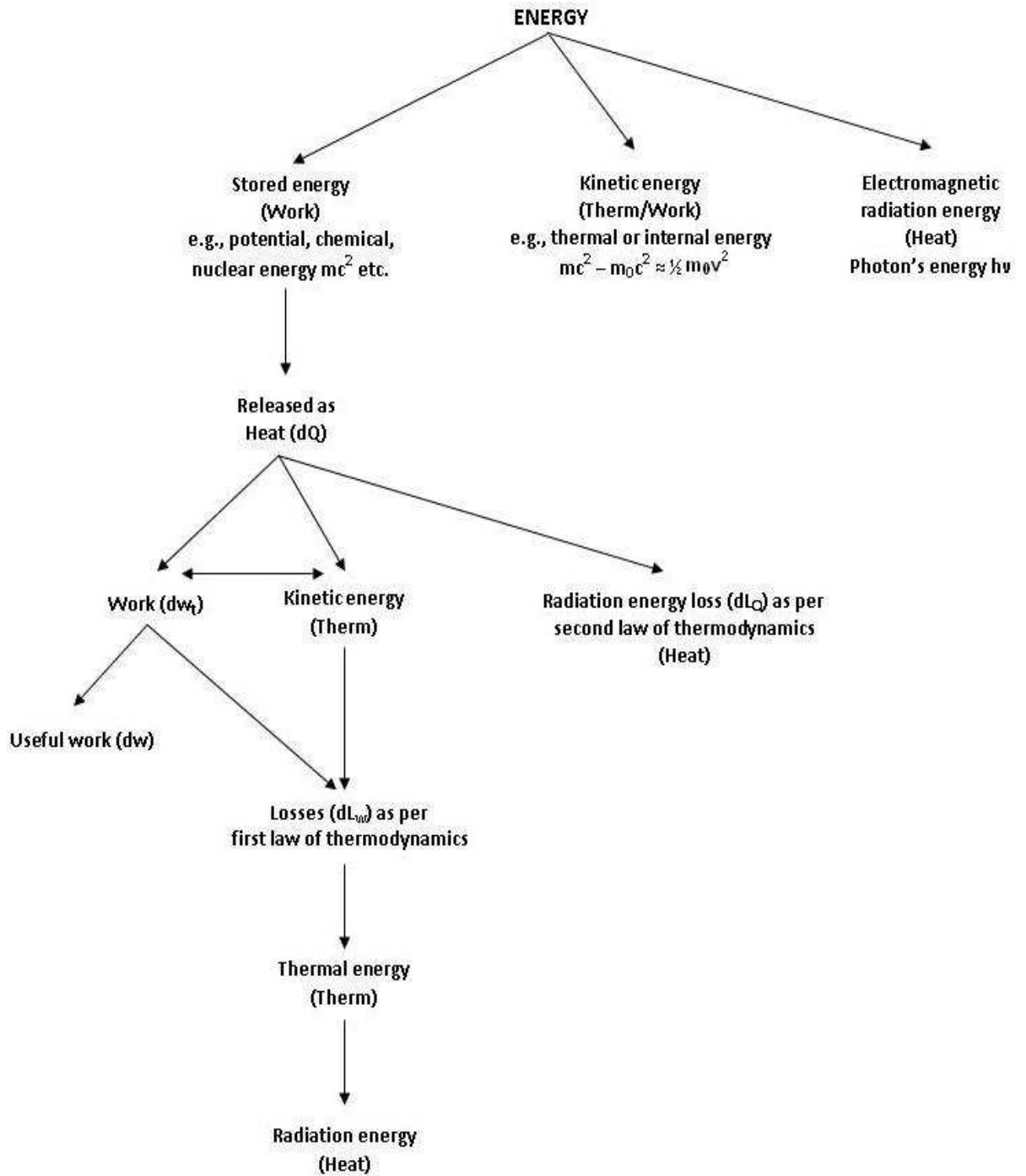}
\textbf{\caption{\textit{\textsl{Another look on what is energy and
its fate.}}}}
\end{figure} To reconcile with the
prevailing/conventional concept of heat-transfer; instead of
categorizing energies into only heat and work, if we categorize (as
follows (Figure 2)) energies into \textit{three} categories namely -
heat, work and therm (the term for the internal or thermal energy),
the whole confusion seems to fade away. It may, however, be seen
that the complex inter-conversion among work, therm (thermal-energy)
and heat occurs. For example, the fuel's chemical energy is first
released as heat (dQ) which converts to thermal-energy (dU) and work
(dW) plus losses (Figure 2).\\\\
Work (dW) is work and heat (dQ) is
heat, but the thermal energy (dU) i.e., therm (though actually being
manifestation of stored work) is considered as heat; since heat $
(h\nu) $ and thermal-energy $ (\frac{1}{2}m_{0}v^{2}) $ both gives a
feel of temperature T, as explained earlier.\\\\
So in the new light of work, therm, and heat; the modes of heat 
transfer conduction and convection is through therm-transfer whereas 
radiation is due to photon-transfer. You feel the heat of a coke-oven 
at a distance through photon-transfer, but if you touch the hot-coke 
the feel of heat (hotness) is due to therm transfer. If we consider 
the `therm and photon' together as one word `caloric', the confusion 
about heat transfer fades away in favor of caloric-transfer, but this 
reminds and revives the old `caloric theory' [13-15] of heat-transfer.
\section{Discussions}
`Law(s) of thermodynamics' are not simply meant for engineers only. Its 
great- ness goes much beyond \cite{ref23,ref24}. It is said \cite{ref24} 
`that the four laws of thermody-namics (Zeroth, First, Second and
Third) drives the universe, and that not knowing (appreciating) the second-law 
of thermodynamics is like never having read a work of Shakespeare'! Physicists 
have great respect for special-relativity and other physical-laws, but 
in-general ignore the importance of the second-law of thermodynamics.\\\\
The first law of thermodynamics tells about \textit{equivalence} of
work and heat, energy-wise, indicating energy-conservation. But the
second law of thermodynamics tells about \textit{non-equivalence} of
work-to-heat and heat-to-work conversion-processes indicating
irreversibility (asymmetry); this asymmetry in the second-law could
be due to not-so-obvious but hidden asymmetry in special-relativity.\\\\
The second law of thermodynamics is not simply a law of
thermodynamics dealing with engines and refrigerators. It has much
more significance at fundamental level especially in Physics as
discussed in earlier-papers [2-4]. It is also seen as law of
entropy-increase of the system. It also indicates and establish the
fact of irreversibility. This thus points towards the thermodynamic
arrow of time \cite{ref25,ref26} which differentiates past from
future. The asymmetry hidden in it due to the irreversibility seems
to be the cause of homo-chirality [27-31] in biological molecules.
Thus the key to life and our existence (or anthropic-principle
\cite{ref32}) could be embedded in the second law of thermodynamics.
If the second law of thermodynamics is so important fundamentally
and that it differentiates between `work to heat' or `heat to work'
conversion, it would be right-time to clearly elaborate and re-define
heat and work in the right perspective to avoid/remove any
misnomer/misconception unfortunately prevailing till now.\\
\section{Conclusions}
Heat and work both are energy, but it is not obvious what the
different forms of energies are: work or heat. From the
right-perspective of second-law-of-thermodynamics it is concluded
that `heat is carried by mass-less messenger particles photons,
whereas work is carried by mass-ive material particles fermions'.
The prevailing misnomer/misconception about heat and heat-transfer
is removed. Studies on `single photon interaction' indicates that
though in-general heat is a statistical-property but there exists
heat property and temperature to single photon too. To reconcile with
the present understanding of heat-transfer, a new term `therm' is
used for internal-energy, the `therm and photon together' reminds of
the old concept of `caloric'. Interestingly, the different fields of
study `thermodynamics' and `relativity' are seen to be interlinked.\\
\section*{Acknowledgements}
The authors thank to V. P. Gautam, Amitabh Ghosh, M. S. Kalara, V. K. Jain,
V. B. Johri, M. M. Verma, G. P. Gupta for their comments and advice.
Institution IET, UPTU Lucknow and GLAITM, Mathura are also thanked
for direct/indirect support. One of the authors (A. Pradhan) thanks the 
Institute of Mathematical Sciences (IMSc.), Chennai, India for providing 
facility under associateship scheme where part of this work was carried out.

\end{document}